# EngThrive: Make It Fast and Easy to Do Great Work

Brian Houck, Tim Bozarth, David Liu, Dean Carignan

## 1. Introduction

Every engineering leader is asking the same questions: "Are my developers productive?" and "Is our productivity improving?". These questions have mattered for decades, but the rise of AI makes them existentially urgent by fundamentally changing how work gets done. This paper proposes a durable, multi-dimensional model for measuring productivity that remains meaningful as the shape of software development continues to evolve.

This paper introduces **Engineering Thrive** (EngThrive), a measurement and improvement system developed at Microsoft. EngThrive's telemetry platform and survey program covers tens of thousands of Microsoft's developers. EngThrive operationalizes multi-dimensional productivity measurement into three dimensions: **Speed**, **Ease**, and **Quality**, with **Thriving** as a guardrail to ensure developer experience improves alongside productivity. We describe the framework's design principles, its metrics, how the metrics were chosen, the supporting measurement infrastructure, and several case studies demonstrating its use. We also show how EngThrive functions as a general-purpose evaluation language that is applicable not only to developer tools and AI, but to organizational design, workplace policy, and the wide array of factors that shape how engineers experience their work

Our work builds on a core insight from the study of productivity in the past decade: Productivity (unlike job performance) cannot be measured discretely on an individual basis. The industry has long sought a simple, reductive way to assess every individual's productivity, but this approach does not work. Productivity reflects both individual capability and the system within which people operate. EngThrive focuses on creating a clear and actionable way to measure the productivity of a system, identify bottlenecks, drive continuous improvement, and lift the effectiveness of individuals within that system.

EngThrive also addresses a fundamental measurement problem: Organizations often track activities (lines of code, pull requests, tasks) and treat them as proxies for outcomes (value delivered, speed, quality). These are not the same. Conflating them produces systems that are precise but wrong, leading to metric gaming and unintended behaviors that move organizations away from desired outcomes.

Unless otherwise noted, findings in this paper are drawn from internal telemetry and survey research conducted across Microsoft's engineering population over several years. This includes repository and build data, collaboration signals, incident data, and large-scale developer experience surveys. While full methodological treatment of each finding is beyond the scope of this paper, individual metrics and their validation are described in the sections that follow.

Our goal is to provide a proven, scalable approach for measuring and improving developer productivity, grounded in our experience with EngThrive at Microsoft. Anchored in Speed, Ease, and Quality, we show how to move from framework to operating system, offering a concrete, citable reference point for organizations attempting to evolve their own systems of work.

## 2. Why Measurement Fails

Before describing what we built, it is worth understanding what we were trying to avoid. The history of developer productivity measurement is littered with well-intentioned efforts that produced misleading results, not because the data was wrong, but because the interpretation was incomplete.

**The COVID paradox.** During the first two months of mandatory remote work at Microsoft in early 2020, pull requests per developer increased by more than 20%. Stock price rose over 15%. By any activity or business metric, things looked excellent. But 78% of developers reported feeling burned out during the same period [7]. Three signals from the same quarter, pointing in two different directions. Any one of them, taken alone, would have told a confident and completely misleading story.

This is not an edge case. It is the norm. Productivity signals routinely diverge, and the choice of which signals to focus on is itself a signal to the organization about what matters.

**Software Engineering is more than Coding.** Studies have shown that AI coding assistants can reduce the time to complete certain coding tasks by up to 56% [5]. It is, however, misleading. Engineering leaders hear “56% faster” and extrapolate to “We should be shipping 50% more innovation.” This extrapolation rests on two misunderstandings. First, leaders dramatically overestimate how much of a developer’s day is spent writing code. Research consistently shows that only about 15% of the average developer’s day involves writing new code. After including testing and debugging, the figure only rises to 25–30% [3, 6]. Second, developer tasks vary enormously in complexity. AI tools are exceptionally effective at reducing the time spent on small, repetitive tasks: boilerplate, scripting, routine refactoring. They are less helpful for the complex, novel, and nuanced problems that constitute the most valuable engineering work. A 56% reduction on one slice of one category of one part of the day does not generalize in the way that a single number on a slide might suggest.

**The single-metric trap.** The deeper problem is structural: Any single metric can be gamed, misinterpreted, or rendered meaningless by context. Lines of code penalize elegance. Velocity without quality is just churn. Satisfaction without output is complacency. The industry has known this for decades, and yet the demand for a single target "productivity" metric persists. The lesson is not that measurement is impossible. It is that measurement without a framework produces noise that masquerades as signal – and worse, creates false confidence that drives confidently incorrect actions. When organizations act on these misleading signals, they do not just fail to improve productivity: They actively degrade it by optimizing for the wrong outcomes with high conviction.

## 3. From Frameworks to Operations

The intellectual foundations for EngThrive draw on multiple lines of research from the last few years, all of which suggest that developer productivity is irreducibly multi-dimensional. The SPACE framework [1] proposed five dimensions (Satisfaction, Performance, Activity, Communication, and Efficiency) and argued that no single metric could capture the full picture. Subsequent work on developer experience [2] quantified the statistical relationships between factors like flow state, feedback loops, and cognitive load and outcomes at the individual, team, and organizational level. More recently, research into the impact of AI on developer workflows [3] confirmed that even the most transformative new tools produce effects that span multiple dimensions and defy simple summarization.

The SPACE framework debunked persistent myths that productivity is purely about activity, that it is only about individuals, that one metric suffices, and it proposed that any credible measurement approach should cover at least three dimensions, include at least one perceptual measure, and expect that well-chosen metrics will naturally create productive tension with one another. These are all core principles that EngThrive puts into action.

DevEx in Action [2] grounded these principles empirically. It proved that improving developer *experience* led to improved individual productivity and creativity, team code quality and technical debt, and organizational retention and profitability. This research quantified how important investments like EngThrive are to an organization's success.

DORA [11] contributed one of the most common measurement frameworks in the industry: Four metrics (Deployment Frequency, Lead Time for Changes, Change Failure Rate, and Mean Time to Recovery) that rigorously capture software delivery performance, including both velocity and stability. These metrics have given countless organizations their first empirical foothold in understanding how well their engineering systems work. EngThrive builds on rather than replaces this foundation. Where DORA measures the health of the delivery pipeline, EngThrive extends the lens to include the human factors, the experience of the developer, and the broader organizational context that shape whether strong delivery performance translates into sustainable productivity.

These frameworks tell you *what* to think about. EngThrive is an attempt to answer the next question: N*ow what do I actually do?*

## 4. The EngThrive Framework

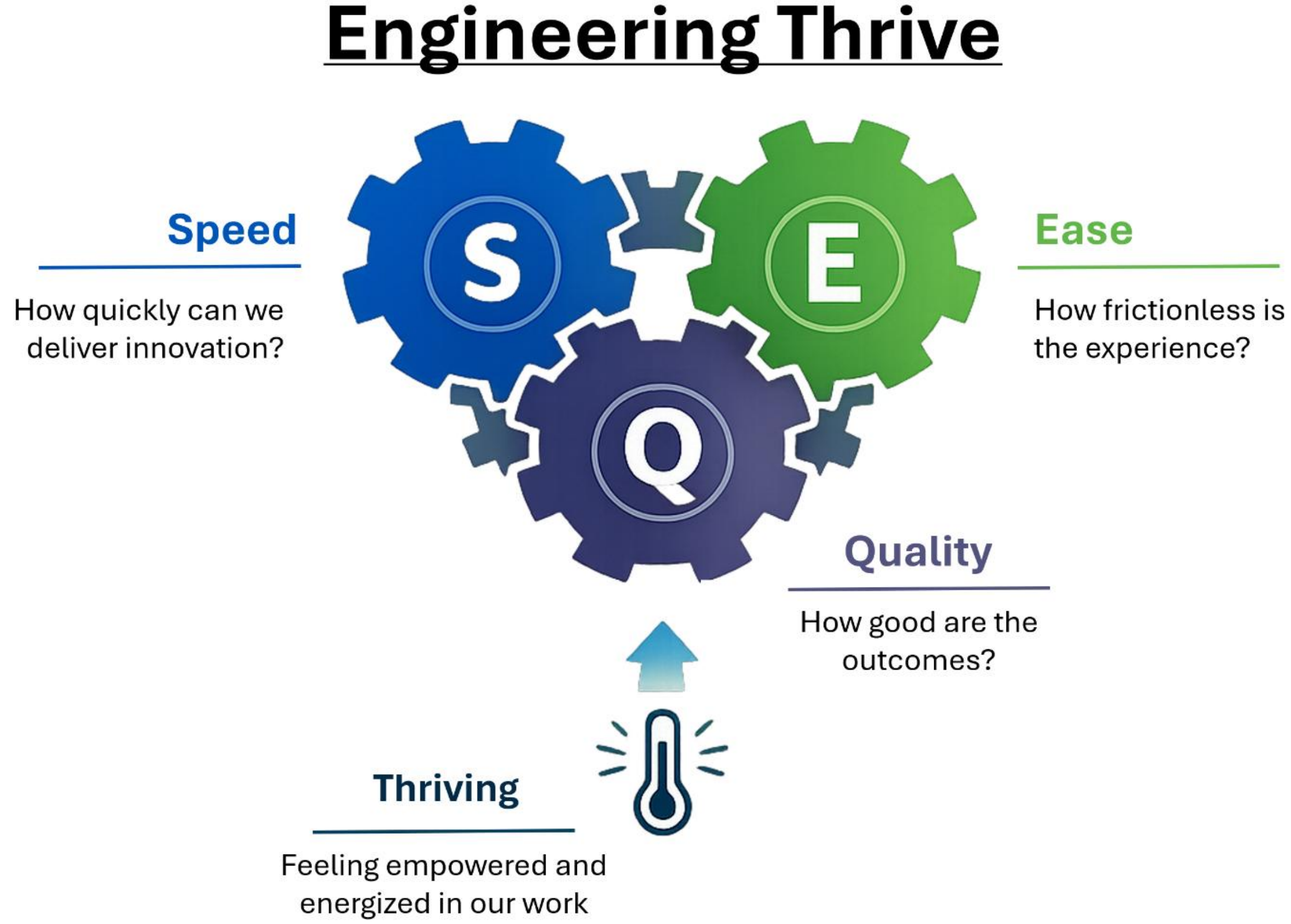

The key objective of EngThrive is to make it fast and easy to do great work. To capture this, the EngThrive metrics are organized around three measurement dimensions – Speed, Ease, and Quality – with Thriving as a guardrail to ensure subjective continuous improvement. Before describing each, we outline the design principles that guided their selection.

Within each dimension, we identify 1-2 North Star metrics. These are our outcome-oriented measures that capture the thing you ultimately care about, and are supported by submetrics that provide diagnostic detail and actionable leverage. The North Stars answer, "A*re we getting better?*" The submetrics enable you to dig into the specific activities and components of the North Star to understand "W*hy or why not?*"

Practically, North Star metrics often require mature, cross-system telemetry that most organizations do not have on day one. The submetrics are where you start — but they are never where you steer. Organizations should define their North Stars early, even before they can measure them, and build toward them iteratively. The danger is managing toward movement in activity submetrics without an outcome metric to anchor them; that path optimizes for motion, not progress.

## 4.1 Design Principles

**Measures vs. Metrics.** We draw a deliberate distinction between *measures* and *metrics*. A measure describes reality: It is a data point, an observation, a fact about the world. A metric decides what matters: It is a measure that has been selected, contextualized, and given a target because an organization has determined it reflects something worth optimizing for. Not every measure should be a metric. The act of promoting a measure to a metric is a statement of values, and should be treated thoughtfully.

**Do No Harm.** Speed, Ease, and Quality are defined as a triad: Each metric exists in the context of the others and is designed to be mutually reinforcing. A metric is only considered improved if it preserves or enhances the other two, ensuring gains are additive rather than a reallocation or concealment of cost. This principle anchors the system: Productivity advances when the triad moves together, not when one dimension improves at the expense of the whole. This principle is more than theoretical; in Section 6.3, we describe an intervention that appeared costly by Speed metrics but was actually free and was enormously valuable by Thriving metrics. Without the triad, it would have been judged a failure or never attempted.

**Gaming alignment.** A common objection to productivity metrics is that they can be gamed. We treat this not as a flaw to be prevented, but as a design constraint to be embraced. Well-designed metrics align gaming behavior with meaningful outcomes. Consider Time-to-First-PR for new hires. When one organization deliberately "gamed" this metric by assigning trivial first-day PRs, the gaming itself produced lasting positive outcomes. Not despite being gamed, but because the act of gaming forced exactly the right behaviors (Section 6.2). The goal is to choose metrics where gaming is the improvement, not simply change in activity.

**Mixed methods.** Every level of EngThrive measurement combines objective telemetry (system logs, repository data, calendar signals) with subjective survey data (developer self-reports on satisfaction, barriers, and perceived experience). Neither alone is sufficient. Telemetry can show that a developer’s focus time dropped; it cannot explain why. Surveys can reveal that developers feel frustrated by their deployment pipeline; they cannot quantify the magnitude of the problem across tens of thousands of engineers. The combination is more than additive. It is the difference between seeing a symptom and understanding a condition.

## 4.2 Speed: How Quickly Can Developers Deliver Innovation?

Speed captures the rate at which developers and teams convert intent into delivered work – measured primarily with the unit of “elapsed calendar time”. Leaders and developers both understand speed intuitively, and the metrics in this dimension are designed to be immediately interpretable.

**Primary metric: Idea-to-Customer (I2C).** I2C measures the elapsed time from the first planning constructs to the moment that work is in the hands of customers. It is the north star for Speed because it captures the full pipeline, not just one stage of it. A team can have fast PR velocity and still be slow if requirements are unclear, reviews are bottlenecked, or deployments are gated behind manual approval processes. This is precisely why PR velocity is an interesting activity measure but is an inadequate proxy for Speed on its own. It captures throughput at one stage of the process but says nothing about the time spent before the first commit or after the last merge. I2C sees all of that. It answers the question leaders actually care about: *How long does it take to turn intent into impact?*

Today, the artifacts that developers produce are code-first, and many of our Speed metrics reflect that reality. We expect this to evolve. As AI-assisted development matures, the workflow will increasingly become spec-driven and intent-driven, with code as a generated artifact rather than the primary unit of authorship. As that shift happens, the meaning of a pull request will change, but I2C will remain stable, rooted in the end-to-end outcome rather than any single intermediate artifact.

**Key Submetrics**

*Time to Nth PR* measures how quickly new engineers begin contributing to their team's codebase and reach certain check-in milestones. Our targets are intentionally aggressive: first PR within seven days, tenth PR within the next few weeks. By a new hire's tenth pull request, we have a greater than 50% chance to predict their coding output patterns over the following two years. This is not merely a speed indicator; it is a leading indicator of long-term engineering trajectory.

What makes this metric particularly interesting is that the content of early PRs matters far less than the fact that they happen. Even trivial first contributions like fixing a typo or updating a configuration file produce positive long-term outcomes. The code is not the point; the act of submitting a PR forces a developer to set up their environment, navigate the codebase, learn the team's review conventions, and ship something. Each of these is independently valuable. We explore this dynamic further in Section 6.2.

*PR Completion Time* measures the elapsed time from PR creation to merge. This is the developer's experience of the review loop, the feedback cycle that determines how quickly work moves from "done on my machine" to "merged and delivering value." Our analysis of the PR lifecycle at scale confirmed a tight coupling: Reducing completion time does not merely feel better; it produces measurably higher throughput across the system. PR Completion Time is one of the most directly actionable submetrics within I2C, because it surfaces bottlenecks that teams can address locally.

## 4.3 Ease: How Frictionless Is the Experience?

Ease captures the degree to which developers can do their work without unnecessary friction, interruption, or toil. If Speed measures the rate of forward progress, Ease measures what is getting in the way, and these metrics are primarily rooted in units of “developer time.”

**Primary metric: Innovation Time Ratio**. Innovation time is a measure of the number of hours in an average period (e.g., a week) that we have to spend on creating value, vs on "run the business” or “business overhead” tasks.

Innovation work includes activities that create new value: building product features, improving user experiences, investing in automation that eliminates classes of toil, and enhancing test and release pipelines to improve quality. “Run the business” work keeps the system operating: migrations, on-call operations, incident response, diagnosing and fixing bugs, and maintaining existing services. Business overhead captures coordination and administrative load: recurring status meetings, reporting, compliance and security tasks, process overhead, and other work required to operate the organization but not directly tied to creating or sustaining product value.

We estimate Innovation Time by starting with Focus Time and using telemetry to identify development activities that erode innovation time (e.g., incident work, build failures, meeting load, compliance tasks) and subtract them from total focus time. This yields a directional, system-level measure of protected time for value creation.

We complement the telemetry with simple surveys (e.g., “What percentage of your week is spent on innovation, running the business, and overhead?”). Even with small sample sizes, these surveys are highly actionable and help validate and calibrate telemetry-based estimates at the team level.

**Key Submetrics**

*Perceived Ease of Delivery* captures, via survey, how difficult developers find it to get their work done within the current system. This is the subjective complement to Innovation Time Ratio. Telemetry can show that a developer lost hours to meetings or build failures, but only the developer can tell you whether the remaining hours felt productive or were spent navigating unclear requirements, waiting on approvals, or context-switching between unrelated tasks. When Perceived Ease diverges from Innovation Time Ratio it surfaces systemic problems that telemetry alone cannot detect.

*Anti-Innovation Time Factors* decompose the telemetry side of the equation. We track meeting load, incident response hours, time spent on compliance and security obligations as distinct erosion categories. Each is independently measurable and actionable. The decomposition is what makes Innovation Time Ratio diagnostic rather than merely descriptive.

## 4.4 Quality: How Good Are the Outcomes?

Quality captures whether the work being done is producing durable, reliable results. Speed without quality is just churn; a team that ships fast but breaks things constantly is not productive, it is generating rework. Quality itself is multi-faceted, it encompasses the defects that escape to production, the speed of recovery when things go wrong, and ultimately, customer satisfaction with what was delivered. The metrics we have chosen, focus on the dimensions most directly tied to developer experience and sustainable engineering, rather than attempting to capture every aspect of quality in a single measure.

We use two complementary North Stars in this dimension. One captures the frequency of quality-related disruptions (PRs-Per-Incident), the other captures the cost of those disruptions (Incident Mitigation Time).

**Primary metric: PRs-Per-Incident.** PRs-Per-Incident captures the ratio of forward progress (pull requests completed) to operational disruption (incidents filed). It is, in effect, our framing of change failure rate: Rather than measuring the percentage of deployments that cause failures, we measure how much forward-looking work a team completes relative to the disruptions it generates. Higher is better. A team completing twenty PRs for every incident is in a fundamentally different position than one completing three.

PRs-Per-Incident is a north star for Quality because it captures the balance that matters most: Are we building faster without breaking more? This metric is deliberately two-sided. A low ratio could indicate either too many incidents (a quality problem) or too few PRs (a velocity problem). The diagnostic value is in the ratio, not in either numerator or denominator alone. This makes it a natural complement to I2C on the Speed side; together, they answer the question of whether a team is delivering innovation sustainably.

**Primary Metric: *Incident Mitigation Time*.** Incident Mitigation Time measures the elapsed time from incident detection to mitigation for live-site issues. It is, for all intents and purposes, our version of the standard industry MTTM metric. Importantly, this metric is framed as a measure of *disruption to the developer experience*, not customer-facing quality per se. When a live-site incident occurs, it pulls developers out of their planned work, destroys focus time, and causes dissatisfaction. Faster mitigation means less disruption to the engineering system as a whole. Incident Mitigation Time tells you not just whether quality is high, but how much it costs the team when quality lapses.

## 4.5 Thriving: The Guardrail

Thriving is not a fourth dimension alongside Speed, Ease, and Quality. It is a guardrail, a check that ensures our optimization of S/E/Q do not sacrifice the people doing the work. The metrics in this area are measured in units of developer happiness.

**Primary Metric: NSAT.** We measure sentiment through two complementary instruments. The first is NSAT (Net Satisfaction), derived from our engineering experience survey, which captures developers' satisfaction with the engineering system: their tools, processes, and workflows. The second is a broader Thriving score drawn from Microsoft's employee sentiment program [10], which measures whether employees feel empowered to do their best work, energized by what they do, and that their work is meaningful.

The evidence for why this guardrail matters is unambiguous. Developers who are unhappy in their jobs are 25 times more likely to report being unproductive and twice as likely to leave within a year [4]. This is not a "soft" concern. It is a direct predictor of the capacity and continuity of an engineering organization. We treat Thriving pragmatically: If an intervention improves Speed, Ease, and Quality metrics but Thriving declines, something is wrong with the intervention. The guardrail catches it and enables us to examine where our understanding of the bottlenecks and lived developer experience is incomplete or incorrect.

**Primary metric: Bad Developer Days (BDD).** Where NSAT captures how developers feel about their experience, BDD captures what actually happened to them. BDD is a composite metric, extending ideas first developed at Google, that quantifies daily developer toil. While the concept of BDD is universally applicable, the specific sub-metrics contained within BDD are distinct to each individual business.

A "bad day" is one in which a developer encounters a threshold-triggering amount of friction from any combination of the identified submetrics: excessive context switching, incident response and live-site toil, build failures or slowness, insufficient focus time, or compliance and security obligations.

BDD is the north star for Thriving because it rolls up many forms of developer frustration into a single, comparable number that can be collected in real-time. When we computed BDD across the organization, the result was sobering: 52% of all developer-days qualified as "bad." Developers experiencing three or more bad days per week were three times more likely to quit their jobs and produced 20% less code than their peers with fewer bad days. BDD is, in essence, a toil tax, and as we have argued elsewhere, toil is a tax on innovation [8]. Every hour a developer spends fighting flaky builds or recovering from unnecessary context switches is an hour not spent on the creative, high-value work that organizations actually need.

For BDD to stay actionable and informative it must adhere to the following criteria: The number of metrics included in the composite must be small (experimentation has shown that more than 6 metrics total adds confusion), and each metric must be independently measurable on the same timeframe as the others (i.e., daily). These conditions are essential to ensure the composite metric does not become overly abstract

The framing of Thriving through BDD is intentionally pragmatic. We are not asking whether developers are delighted. We are asking whether they can do their jobs without the system working against them. The distinction matters: Delight is an outcome of removing friction, not a goal unto itself.

## 5. The Measurement System

A framework without infrastructure is an aspiration. To make EngThrive a reality within Microsoft, we built a complete measurement system designed to make multi-dimensional data accessible, actionable, and safe.

### 5.1 The Data Platform

Before dashboards, surveys, or analyses could be useful, we needed a foundation: A unified data platform that pulls developer telemetry from dozens of disparate systems into a single, consistent, normalized layer. Source control data from GitHub, build telemetry from internal CI/CD systems, calendar signals from Outlook and Viva, incident management data, compliance tracking, and more are all joined and reconciled into a common data model. This platform is the bedrock for everything that follows. Metrics are computed consistently across organizations, survey responses can be enriched with telemetry for deeper analysis, and researchers can study cross-system relationships (like the connection between meeting load and PR velocity) that would be impossible to detect in any single data silo. Without this investment, the mixed-methods approach that EngThrive depends on simply would not work at scale.

It is also important to remember that this system did not appear as a comprehensive data platform in a day. It was done iteratively, with each step prioritized to enable us to materialize a more comprehensive image of the EngThrive North Star metrics.

### 5.2 The Survey

The EngThrive Engineering Experience Survey is administered across Microsoft's global engineering population. It is structured around the Speed, Ease, and Quality dimensions, with each section containing both Likert-scale items and barrier-identification questions. Developers report which specific barriers are most impacting their experience (deployment friction, build reliability, meeting load, unclear requirements) and these barrier responses are linked directly to satisfaction scores to identify where interventions will have the highest impact.

The survey also captures time allocation: Developers estimate what fraction of their week is spent on innovation (new features, improvements), running the business (maintenance, operational work), and overhead (meetings, compliance, administrative tasks). This decomposition is analytically powerful because it surfaces structural problems. An organization where developers spend 30% of their time on overhead has a fundamentally different challenge than one where they spend 30% on operational toil, even if both report the same overall satisfaction.

AI-related questions are woven into the existing dimensional structure rather than isolated in a separate section. This is a deliberate design choice: AI is a tool that affects Speed, Ease, and Quality, not a separate dimension of experience to be measured independently.

## 5.3 Dashboards and Personas

EngThrive data is surfaced through a dashboard ecosystem designed for different personas. **Organization dashboards** serve leaders of groups with 50 or more software engineer individual contributors; they emphasize trends, cross-team comparisons, and strategic resource allocation. **Team dashboards** serve managers and tech leads of groups with five or more members; they emphasize barrier identification and local improvement opportunities.

Absent from this hierarchy is an individual-level dashboard – by design. EngThrive metrics are not individual performance measures and are never used to evaluate, rank, or compare individual engineers. The system measures the environment in which developers work, not the developers themselves. Managers and leaders are accountable for the environment they create, and the dashboards are designed to reflect that accountability.

Privacy is a first-order design constraint. Organization dashboards are aggregated over larger populations, and the Team dashboard is scoped to the team's own data. Aggregation thresholds prevent identification of individuals in small groups. This is not merely a compliance requirement. It is foundational to trust, and trust is foundational to data quality.

## 5.4 Culture, Insights, Infrastructure

If Speed, Ease, and Quality describe *what* EngThrive measures, the three pillars describe *how* change actually happens. EngThrive operates through three mutually reinforcing pillars.

**Culture** encompasses the organizational norms, leadership behaviors, and shared expectations that determine whether measurement data actually drives change: whether leaders act on what the data reveals, whether teams feel safe surfacing problems, whether improvement is rewarded.

**Insights** encompass the data itself: the surveys, telemetry, analyses, and research that produce understanding.

**Infrastructure** encompasses the tooling, dashboards, and systems that make insights accessible and actionable at scale.

No pillar works alone. Insights without culture produces shelfware: beautiful dashboards that no one acts on. Culture without insights produces intuition: well-intentioned leaders making decisions in the dark. Infrastructure without either produces capability without purpose. The system works when all three are functioning together.

# 6. Case Studies

The following case studies help demonstrate the principles of EngThrive in action, taking you through examples of bringing data to insights to action to impact.

## 6.1 Focus Time in CoreAI

Research has shown that "Too Many Meetings" is the #2 most frequently cited workplace challenge among software engineers across the entire industry [12]. Relatedly, developers who have a significant amount of time carved out for deep work feel 50 percent more productive compared with those lacking in dedicated time [2].

In late 2025, the Microsoft CoreAI organization, which has thousands of developers, launched a dedicated initiative to decrease unnecessary meeting load and increase developer focus time. Executive leadership set an explicit target: Lift the bottom 20th percentile of developers to at least 25 hours of focus time per week. To drive this, they established biweekly leadership reviews, shared tracking, and a cross-team collaboration channel.

Teams identified and implemented five common intervention patterns: reducing calendar fragmentation by clustering collaboration into designated windows; improving meeting hygiene through required agendas and explicit permission to decline; protecting deep-work blocks through calendar tools and "AM sync / PM think" norms; coaching outlier teams using dashboard data; and addressing structural barriers like on-call patterns and unscheduled interruptions.

Within eight weeks, the outcome metrics improved across multiple EngThrive dimensions: Focus Time increased by 2.1 hours per developer per week (2x the control group), and Bad Developer Days declined by 25%. These are the results that matter. To understand why they improved, we look to the supporting activity metrics which changed in ways consistent with the outcome gains. PR velocity rose 13% (about 4x the control), equivalent to the output of ~350 additional developers. But it isn't just about going faster. Only half of this lift in BDD is explained by increased focus time (a component of BDD); the remainder appears to come from teams using their added capacity to reduce technical debt that had been driving bad days. Status meetings decreased by 10% and multitasking in meetings dropped by 8% (a proxy for low-quality meetings), while manager one-on-ones increased, indicating stronger leadership engagement. Taken together, the pattern is clear: Outcome metrics improved, and the activity metrics explain the mechanism. Less low-quality collaboration, more protected focus, and higher-quality interactions, without displacing other dimensions.

This case study illustrates EngThrive's operating model: Validated metrics create visibility, visibility enables leadership attention, leadership attention empowers teams to identify and implement local solutions, and improvements reinforce the metric's value. As one team lead described it: "Metrics led to questions, questions led to improvements, improvements reinforced the metric's value."

## 6.2 Onboarding Speed

One experiment to try and "game" Time-to-First-PR had interesting results. One organization of approximately 4,000 developers invested in accelerating new-hire onboarding by purposefully assigning a trivial first pull request. They found that if they wanted to game the metric, they could. They saw a 30% improvement in Time-to-First-PR. But surprisingly, even gaming the metric led to good long-term outcomes. Those new-hire developers that were part of the experiment had 23% more pull requests over their first twelve months than the control group. Interviews with experiment participants illustrated the underlying cause: The first pull request isn't about the code itself, not to the new hires. It's about setting up the environment. It's about learning the tools and processes. It's about learning the language of the team.

Additionally, an AI-assisted onboarding tool called FirstMate reduced Time-to-First-PR by 65%, an effect driven not by making the first PR easier, but by automating the environment setup, documentation discovery, and codebase orientation that historically delayed new hires for days or weeks.

This case study demonstrates the power of EngThrive's focus on "Outcome" metrics. Even if they are "gamed", the impact of those results still create the outcome of making it faster and easier to build great products.

### 6.3 Health Days

During the 2020 burnout crisis, one organization ran an experiment in which all developers received two unplanned days off, called "Health Days." Pull request output during those days dropped, as expected. But all "lost" pull requests were made up for within 14 days, while the burnout relief persisted for 14 weeks [4]. This is the Thriving guardrail in action: A wellbeing intervention that appeared costly by Speed metrics was actually free by Speed metrics and enormously valuable by Thriving metrics. Without measuring both, the intervention would have been judged a failure or never attempted at all.

## 7. Beyond Developer Tools

EngThrive was built to measure the developer experience. But the experience of a developer is not shaped only by their IDE, their CI/CD pipeline, or their AI coding assistant. It is shaped by every factor that influences their ability to do focused, meaningful, high-quality work, including factors that have nothing to do with software. Not all these factors are within an engineering leader's power to change, but the ability to measure their impact is valuable even when the lever belongs to someone else. A quantified effect on Speed, Ease, Quality or Thriving gives leaders evidence to make the case to the people who can change it.

**Office design.** We have measured that developers who move into more modern office buildings become more collaborative, more engaged, and more productive. The mechanism is not mysterious: Access to natural light and better HVAC systems serve basic human needs. When those needs are met, people perform better. When developers work in legacy buildings with poor ventilation and limited daylight, the engineering system suffers, not because the code is different, but because the humans are different. Internal research has shown that building design accounts for up to a 10% difference in overall developer engagement.

**Weather.** In Seattle, the northernmost major city in the continental United States, developer coding patterns over the winter can reveal which days were snow days, because output drops approximately 18%, according to our internal research. The reasons are confounded (childcare, commute disruption, mood), but the effect is real and measurable through the same Speed metrics used to evaluate AI tools. While weather might not be changeable (like other global health events), understanding its impact may be useful for planning purposes, or even predicting deployment velocity.

**Policy.** Developers who perceive high levels of organizational bureaucracy are 70% more likely to be actively looking for jobs at other companies [12]. Vacation time, which many leaders fear will reduce output, does not reduce it, but it provides meaningful protection against burnout. These are policy questions, not technology questions, but they are evaluable through the same framework.

**Organizational structure.** The ideal PM-to-developer ratio, the appropriate manager span of control, the impact of return-to-office mandates: These are questions about human systems, not engineering systems. But they all affect Speed, Ease, Quality, and Thriving, and EngThrive can measure their impact using the same instruments and the same analytical approach that it uses for everything else.

This universality is not incidental. It is the reason EngThrive is durable. Any framework that measures only the impact of software tools becomes obsolete when those tools change. A framework that measures the human experience of doing engineering work remains relevant regardless of what tools, policies, or organizational structures are in place, because the human needs underlying it do not change.

## 8. AI as Application, Not Exception

Every engineering leader we speak with asks the same question: *How is AI impacting my developers' productivity?* The strength of EngThrive is that the framework remains durable even as AI fundamentally changes how work gets done. This durability comes from its focus on Speed, Ease, and Quality - outcome metrics that span the end-to-end developer experience. AI introduces a powerful new tool that can both remove bottlenecks and reshape how development work happens. While AI offers significant potential for improvement, it should be evaluated using the same framework and metrics applied to any other intervention—such as a new onboarding program or meeting policy. The problems may differ, but the principles for measuring improvement and transformation remain consistent.

That said, the impact of AI is worth examining in detail, as it highlights many of the challenges discussed throughout this paper.

**What developers report.** Among developers who regularly use AI tools, 88% say it increases their task throughput, 82% say it increases their efficiency, 71% say it increases their ability to deliver customer and business value, 62% say it increases their job satisfaction, and 49% say it improves their ability to collaborate effectively [3]. The impact is not even across dimensions, but it is present across all of them. Any measurement approach that captures only one dimension will miss the majority of the effect.

**What the data shows.** Multiple studies have shown that AI use is increasing PR velocity [3,5,9]. This demonstrates a meaningful change in development activity, but is only an activity signal - not an outcome. More pull requests do not inherently mean more value delivered, fewer incidents, or faster innovation. This is precisely where EngThrive's North Star metrics matter: They let us ask whether we are making progress, not just increasing motion. A Microsoft engineering organization of 3,000 developers leaned into adoption of Azure SRE Agent to reduce toil from managing live-site issues. That organization has seen their Incident Mitigation Time, a Quality North Star, improve 2.4X faster than the company as a whole. While this was not a controlled experiment, it provides evidence that agentic AI tools applied to specific problems can produce measurable improvements in outcomes, not just changes in activity.

**What moderates the effect.** The impact of AI is not uniform. Individual usage frequency matters: Developers who use AI daily report substantially higher confidence in its benefits than those who use it occasionally. Team adoption matters: As more peers on a team use AI, each individual developer's perceived productivity gains increase, suggesting a compounding flywheel effect driven by shared learning. And organizational culture matters: Teams that actively advocate for AI adoption see approximately 10% higher confidence in AI's productivity impact; teams that provide formal documentation see about 3% higher confidence [3]. The culture surrounding AI tools may matter as much as the tools themselves.

**Activity vs Outcomes in the Age of AI.** The distinction between Inputs and Outcomes is particularly relevant when thinking about AI. **Activity metrics** capture the conditions under which AI-assisted work occurs: adoption rates, usage frequency, model selection, and token consumption. **Outcome metrics** capture the value delivered: idea-to-customer time, bad developer days, PRs-per-incident. The key insight is that AI introduces new activity metrics, forces reinterpretation of some of our activity artifacts (e.g. a PR authored with AI assistance is different from one written entirely by hand, but both deliver value), but AI leaves outcome metrics stable. Outcomes are the north star that AI should accelerate progress against.

The most important point is this: It does not matter how advanced a tool is. The questions are still human questions. How are people using it? What culture enables them? What friction remains? Is the work better? Are the people okay? These are the same questions EngThrive asks about office buildings and meeting policies. AI is not an exception to the framework. It is the most visible current application.

## 9. Getting Started

For organizations looking to adopt a multi-dimensional measurement approach, we offer practical guidance drawn from our own experience and from working with dozens of engineering organizations.

**Get started even with imperfect data.** Do you have precise, telemetry-driven data for every work item, commit, code review, build, deployment, and feature flag? If so, you are in an uncommonly good position to jump into EngThrive. However, even if you have none of those signals, you can jump into this space quickly via specific, EngThrive targeted surveys and basic telemetry. Although Surveys do not give you real-time telemetry, they do allow you to quickly understand the biggest bottlenecks in your system, and to create an accurate set of measures related to the Speed + Ease + Quality North Star EngThrive metrics.

**Pick one or two metrics per dimension.** For each dimension, select a small number of metrics that balance subjective and objective signals. A speed metric from telemetry paired with a satisfaction question from a survey is more informative than five telemetry metrics alone. Metric discipline matters: Too many metrics dilute focus and make it harder to identify what is actually changing.

**Focus on continuous improvement.** Every metric has a numeric value – however, the culture of EngThrive is focused on continuous improvement. The first step is to understand the data you have, form interventions to drive change, and to measure that change over time.

**Expect gaming, and design for it.** Choose metrics the focus on outcomes. If your metric can be improved by doing the right thing, then "gaming" it is indistinguishable from genuine improvement. If it can only be improved by doing the wrong thing, choose a different metric.

**Iterate.** Metrics are living systems. What matters to your organization today may not be what matters in a year. Revisit your metrics periodically, not to chase trends, but to ensure they still reflect your actual priorities and challenges. If a metric is no longer informative, retire it. If a new challenge emerges that your current metrics do not capture, add one, thoughtfully.

## 10. Conclusion

Developer productivity is multi-dimensional, deeply human, and stubbornly resistant to simple measurements. The research community made significant progress in establishing this through DORA, SPACE and DevEx.

EngThrive represents the next step: from principles to practice, from framework to operating system. By organizing measurements around Speed, Ease, and Quality we provide a structure that is concrete enough to implement, flexible enough to adapt, and durable enough to outlast any particular technology cycle.

The case studies presented here demonstrate that this approach works: Outcome-oriented metrics, combined with leadership attention and team empowerment, produce measurable improvements across multiple dimensions simultaneously. They also demonstrate the dangers of single-dimensional and “activity” oriented measures, that feed unintended side-effects and offer incomplete understanding.

Perhaps most importantly, EngThrive is not only a developer tools measurement system. It is a language for evaluating anything that shapes the developer experience, from AI coding assistants to office HVAC systems, from onboarding programs to organizational bureaucracy. The factors change; the human needs for speed, ease, and quality do not.

We offer EngThrive not as the final word, but as a concrete demonstration of how it is possible to make it faster and easier to do great work (and to measure that it’s working). We invite the community to adopt, adapt, critique, and extend this work.

---

## Acknowledgments

We would like to thank Alex Kyllo and the entire EngThrive team for their significant contributions to building and operating EngThrive. We would also like to thank Jay Parikh, Lindsay Folk, Outi Rentola, our collaborators in the People Analytics team, the EngThrive AI Impact Squad, Caitie McCaffrey and Nicole Forsgren for helping get EngThrive bootstrapped, and all our partners across Microsoft (and beyond).